\UseRawInputEncoding
\documentclass[twocolumn,showpacs,preprintnumbers,amsmath,amssymb]{revtex4}

\usepackage{upgreek} 
\usepackage{amsmath,amsfonts, amssymb,graphicx,amsthm,color}
\usepackage{mathtools}
\usepackage{dcolumn}
\usepackage{bm}
\usepackage{rotating}
\usepackage{units} 
\usepackage{amssymb}
\usepackage{sidecap}
\usepackage{comment}

\begin{document}

\title{1S-3S cw spectroscopy of hydrogen/deuterium atom}

\author{Pauline Yzombard$^{1,3}$, Simon Thomas$^{1,2}$, Lucile Julien$^{1}$, Francois Biraben$^1$, Francois Nez$^1$  
}
\affiliation{$^1$ Laboratoire Kastler Brossel, Sorbonne Université, CNRS, ENS-PSL Research University, Collège de France, 4 place Jussieu, Case 74, 75252 Paris Cedex 05, France}

\affiliation{$^2$(current affiliation): Laboratoire de physique des lasers. Université Sorbonne Paris Nord (USPN) - Institut Galilée, 99, av. J.B. Clément - 93430 Villetaneuse, France}

\affiliation{$^3$ corresponding author:  pauline.yzombard@lkb.upmc.fr}

\begin{abstract}
    We study the 1S-3S two-photon transition of hydrogen in a thermal atomic beam, using a home-made cw laser source at 205~nm. The experimental method is described, leading in 2017 to the measurement of the 1S-3S transition frequency in hydrogen atom with a relative uncertainty of $9 \times 10^{-13}$. This result contributes to the "proton puzzle" resolution but is in disagreement with the ones of some others experiments. We have recently improved our set-up with the aim of carrying out the same measurement in deuterium. With the improved detection system, we have observed a broadened fluorescence signal, superimposed on the narrow signal studied so far, and due to the stray accumulation of atoms in the vacuum chamber. The possible resulting systematic effect is discussed.
\end{abstract}

\maketitle

\section{Introduction}
Our group in Paris studies the two-photon 1S-3S transition of hydrogen in an atomic beam for several decades now. The main problem for such a study is the generation of the 205~nm laser radiation needed for the excitation. In a first step, we performed two successive frequency doublings of a home-made titanium-sapphire laser at 820~nm \cite{Bourzeix1993}, with LBO and BBO crystals inside two enhancement
cavities. To avoid rapid deterioration of the BBO crystal, a modulation of the second cavity was needed, leading to a quasicontinuous
regime where the UV intensity consists of $3~\mu s$
pulses at a frequency of 30 kHz \cite{Bourzeix1996}. During the nineties, we performed with this source the determination of the 1S Lamb shift by comparison of the 1S-3S and 2S-6D transitions frequencies which are in a ratio close to 4:1  \cite{Bourzeix1996,Bourzeix1997}.

In this experiment, the excitation takes place in an enhancement cavity and the two counterpropagating UV beams are collinear with the effusive thermal atomic beam, so that the transit time broadening is reduced. Although the first-order Doppler effect is cancelled, the second-order Doppler effect is the main systematic effect since it shifts the signal by about 150~kHz for an atomic velocity of 3~km/s. 

A way to control this effect was proposed by our group in 1991 \cite{Biraben1991}, and more details about the method will be presented in Sec. \ref{sec4}. It was successfully implemented by our group \cite{Hagel2002}, which allowed us to achieve few years later the first optical frequency measurement of the 1S-3S transition with a frequency comb \cite{Arnoult2010}. The result obtained, with a relative uncertainty of $4.5\times 10^{-12}$, made this frequency the second one accurately known of the optical frequencies in hydrogen. 

The first one is indeed the 1S-2S two-photon transition frequency, studied for a long time by the Garching group \cite{Fischer2004}\cite{Parthey2011}. Due to the metastability of the 2S state, this transition has an extremely narrow linewidth, and its frequency is presently known with an accuracy of $4.2\times 10^{-15}$. Combining this frequency with another optical one allows to determine both the Rydberg constant and the 1S Lamb shift, taking advantage of the approximate $1/n^3$ scaling law of the Lamb shifts and the fact that the deviation from this law is precisely calculated \cite{Czarnecki2005}. 

The main contributions to the Lamb shift are QED radiative corrections, which are now calculated with a very high precision. If one assume that their calculation is correct, one can deduce nuclear size effect from the measurement of the Lamb shift. Until 2010, the value of the proton charge radius was deduced either from atomic hydrogen spectroscopy using this method, or from scattering experiments and its commonly accepted value \cite{CODATA2010} was $r_{\mathrm{p}}=0.8775(51)\ \mathrm{fm}$. However, the measurement of the 2S-2P interval performed in muonic hydrogen by the CREMA collaboration in the Paul Scherrer Institute (Switzerland) led to a ten more precise value of this radius $r_{\mathrm{p}}=0.84087(39)\ \mathrm{fm}$ \cite{Pohl2010}\cite{Antognini2013}, but in disagreement by 7~$\sigma$ with the previous one. During the ten past years, an intense research activity, both theoretical and experimental, has been carried out to solve this discrepancy, which is now known as the ``proton puzzle" \cite{Karr2019}. 

In atomic hydrogen, both one-photon and two-photon transitions have been recently studied leading to new frequency measurements: 2S-4P \cite{Beyer2017} and 1S-3S \cite{Grinin2020} in Garching, 2S-2P in York \cite{Bezginov2019}, 2S-8D in Boulder \cite{Brandt2022}. Some of them are in agreement with the measurement of the $r_{\mathrm{p}}$ value deduced from muonic hydrogen spectroscopy, but others do not agree.

While the Garching group uses picosecond pulses to excite the 1S-3S two-photon transition, we have developed in Paris a continuous laser source at 205~nm \cite{Galtier2014} to perform the 1S-3S excitation and carry out a new precise measurement of the transition frequency \cite{Fleurbaey2018}. The use of two different techniques by the Garching and the Paris groups to study the same transition is interesting since it could open the way to reveal possible systematic effects responsible for a shift of the line. The experimental device we used for our measurement and recent improvements of our experimental set-up are described in the following.

\section{Experimental set-up and result obtained in hydrogen}

\begin{figure}[h]
\centering
\includegraphics[width=80mm]{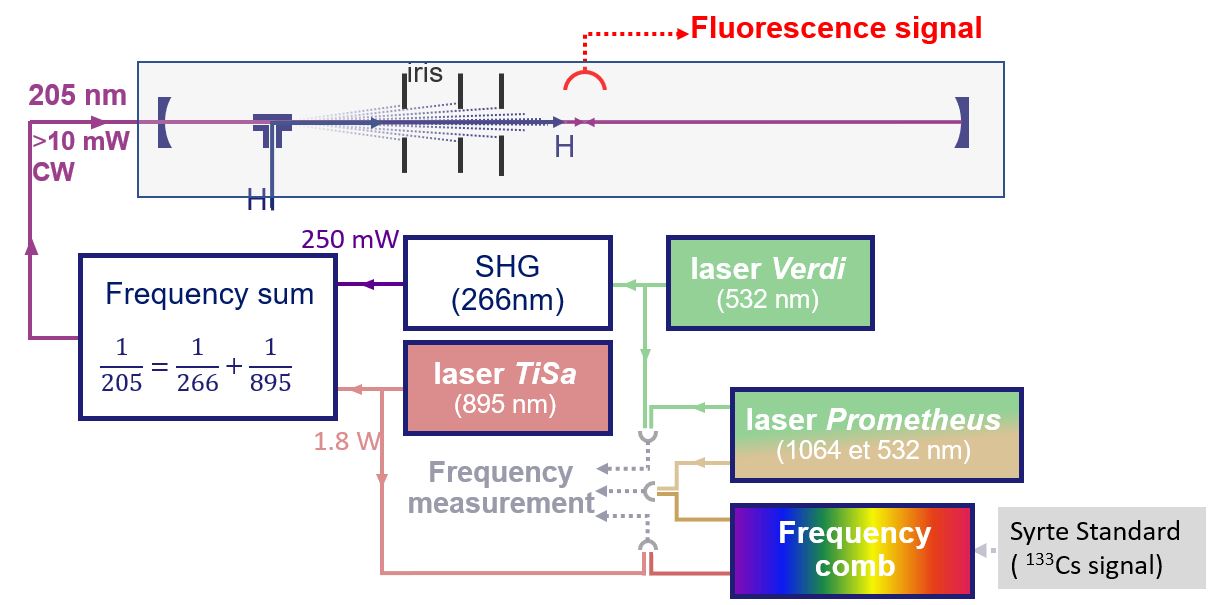}
\caption{Simplified schematic of the experimental setup.}
\label{figSetup}
\end{figure}

In Fig. \ref{figSetup} is shown a simplified schematic of the
experimental apparatus used until measurement campaign of 2020. The hydrogen atomic beam is generated by a MHz dissociator applied on H$_2$ gas that generates about $10^{18}$ atoms/s, injected into the vacuum chamber. A set of irises ensures the collimation of the atomic beam. The two-photon 1S-3S laser excitation is realised thanks to a build-up in-vacuum 205~nm Fabry Perot (FP) cavity. The excitation probability is the highest at the focal point of the FP cavity, on top of which are installed the collective optics and PhotoMultiplier (PM) which record the fluorescence red photons emitted by the decay of the 3S $\rightarrow$ 2P states. A more detailed schematic of the experimental chamber is displayed in Fig. \ref{fig2exp} a). 

To generate the 205~nm cw laser, we use a frequency sum inside $\beta$-barium borate (BBO) crystal, that adds up the radiations of the fourth harmonic of the Nd-YAG at 266 nm,  (resulting from the frequency doubling of the 532 nm {\it Verdi V6 Coherent laser} ) and the home-made tunable titanium:sapphire (Ti:Sa) laser at 894 nm \cite{Galtier2014}. Typically 12mW of 205~nm can be injected into the Fabry-Perot cavity. 
The frequency stability of the lasers used to be ensured via  several Fabry-Perot cavities, all referenced to the radiation of a laser diode at 778 nm locked on a two-photon hyperfine transition of $^{85}$Rb \cite{Touahri1997,Galtier2015} . Recent improvements on the locking system of the laser system have been realised during the PhD thesis of S. Thomas \cite{PhD_Thomas}, which have led to the possibility of scanning the laser frequency in a wider range and will be discussed in the following section \ref{sec3}. In order to measure the different laser frequencies, we use a {\it Menlo Systems} femtosecond (fs) frequency comb, whose 780 nm output is spectrally broadened in a photonic crystal fiber (PCF), to beat each of our lasers with one of the teeth of the combs and counts the resulting RF signal. 
Despite the PCF, the broadening of the fs-comb signal is quite limited around 532~nm, making the beatnote between the VERDI and the fs-comb too weak to reliably extract its frequency.  Thus, since 2016, 
we use an intermediate Nd-YAG laser ({\it Prometheus, Innolight}), providing both 1064~nm and 532~nm outputs. The 532~nm beam is beaten with the VERDI laser, whereas the 1064~nm output is compared to the fs-comb \cite{PhD_Fleurbaey}. Finally, the frequency comb is referenced to the LNE-SYRTE Cs fountain primary frequency standards
via fiber link \cite{Touahri1997}.

\section{Recent improvements}
\label{sec3}
Two main improvements have been implemented in our experimental setup, over the past years.
First, to improve the statistics, we worked on reducing the background signal and on improving the general stability of the 205~nm Fabry-Perot locking system.  

\begin{figure}[h]
\centering
\includegraphics[width=80mm]{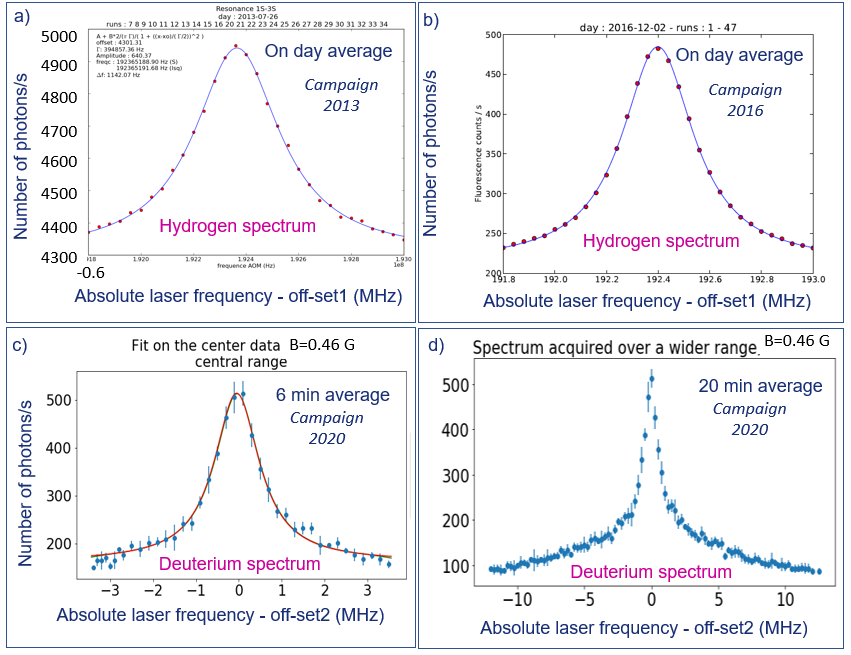}
\caption{typical 1S-3S fluorescence signals during hydrogen {\it a.) and b).} and deuterium {\it c.) and d).} campaigns. Several technical improvements led to the background reduction and to a wider scanning laser frequency range obtained in d). a) Data taken during S.~Galtier PhD thesis \cite{PhD_Galtier}, b) Data taken during H.~Fleurbaey PhD thesis \cite{PhD_Fleurbaey}, c) and d) Data taken during S.~Thomas PhD thesis \cite{PhD_Thomas}}
\label{fig1fluo}
\end{figure}

\begin{figure}[h]
\centering
\includegraphics[width=80mm]{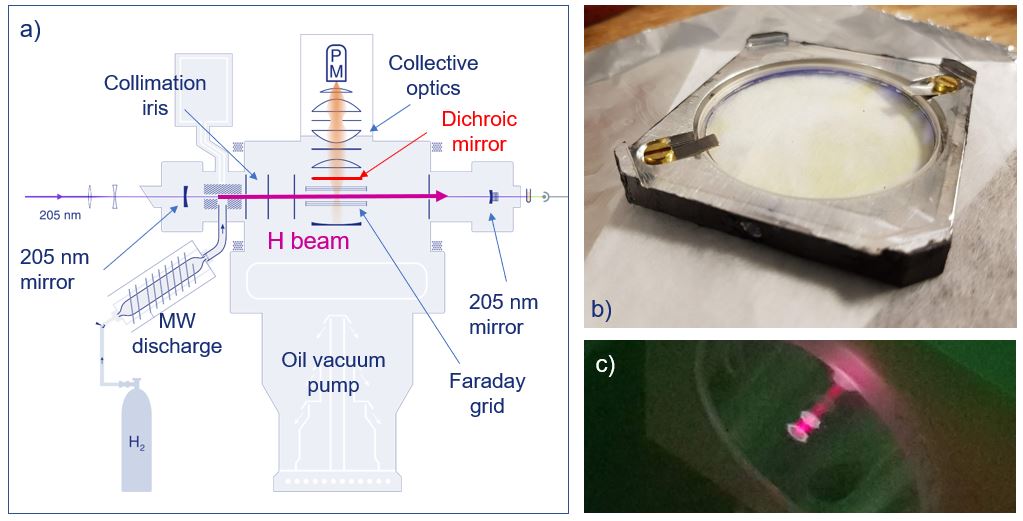}
\caption{a) An overview of the experimental set-up. A dichroic mirror has been recently added to block red fluorescence photons from UV fluorescence into glass ; drastically reducing the background signals. b) The dichroic mirror blocking UV 205~nm photons and allowing visible photons from the $n=3 \to n=2$ de-excitations to be transmitted toward the PMT. c) A typical fluorescence signal coming from UV photons (266~nm beam) fluorescing into glass.}
\label{fig2exp}
\end{figure}

Typical signals during the three last measurements campaigns on the 1S-3S transition in hydrogen and deuterium, are shown in Fig.\ref{fig1fluo}. In a), we can see that the background signal was about 4300~counts/s.  The counting photon electronic module has been changed, and in parallel, we pursued a better calibration for the triggering threshold to counts the photons. We performed also a better isolation of the parasitic photons : in addition to the interferential filter that lets only red photons passing through the PMT, we recently added a dichroic mirror right before the collective optics to block any UV photons to penetrate (shown in Fig.\ref{fig2exp} a) and b)). Indeed, we observed that the UV at 205~nm tends to induce a red fluorescence into glass, as shown in Fig.\ref{fig2exp} c).  These changes in our detection set-up led to a drastic reduction of the background counts signals, to about ~200~counts/s, as we can see in Fig.\ref{fig1fluo} b), c) and d).

The second main improvement concerns the laser system. We worked to increase the frequency stability and tunability of the 205~nm laser. Before 2018, the usual laser frequency scan was about of ±~1.5 MHz around the resonance frequency (meaning a total width scan of the atomic transition of $\pm$ 3~MHz, since the excitation is done via 2-photon absorption). This corresponds to a tunability range of about 5 times the natural linewidth of the transition. It takes about 6~min to acquire a full run, composed of 10 scans of the transition. Each scan is made of 31 data points. It was possible to extend the range to a scan width of ± 5 MHz corresponding to about 10 natural linewidths, and this took about 11~min for 10~scans of 51~points each \cite{PhD_Fleurbaey}. The new locking system of the lasers allows a wider scan up to ±~30 MHz, corresponding to about 60 natural linewidths, and it takes 20~min to record 10 scans. Technical details are available in \cite{PhD_Thomas}.

Quickly, this wider scanning frequency range has been made possible thanks to the design of a new locking scheme mechanism for the lasers. In the past, we used to shift the frequency of the 205~nm beam, playing on the 984~nm frequency using a double-pass AOM to add the desired shift, and keeping constant the frequency of the 266~nm beam. The limitation of this scheme comes from the AOM efficiency that drops quickly when it is tuned out of its resonance circuit. This limits the range of scanning of about ±~4-5 MHz max. The new locking scheme opts for rather tuning the 532~nm laser, that generates the 266~nm radiation by frequency-doubling. This allows to double the effective scanning range of the 205~nm beam, compared to the one of the 532~nm laser. The green laser used to generated the 266~nm is a VERDI V-6 Coherent, that has the capability of being easily tunable over a wide range with not significant drops in intensity. In order to stabilise its frequency, it is locked via a Pound-Drever-Hall locking scheme \cite{PDHlock} to an auxiliary Fabry-Perot (FPA) cavity. We apply an error signal to move the length of the FPA cavity, which changes its Free Spectral Range (FSR) and then moves the resonance peak on which the VERDI laser is locked to. As a consequence, the frequency of 532~nm laser follows the resonance peak and is shifted accordingly. This allows a wider scanning range up to ±~30 MHz on the 205~nm beam with no significant drops nor instabilities in intensity observed. This led to a new systematic effect that we further discuss in the following section.

\section{Analysis of a potential newly observed systematic effect}
\label{sec4}

We present here data taken during the deuterium 1S-3S campaign of October-December 2020, recorded during the PhD thesis of S. Thomas \cite{PhD_Thomas}.

\subsection{Main systematic effect: the Second Order Doppler effect}
The main systematic effects affecting our data are fully characterized and detailed in \cite{Fleurbaey2018}. It is beyond this article to review them all here. In this section, we aim to present the preliminary results in the investigation of a new systematic effect we have recently discovered thanks to the experimental improvements in both signal detection and laser stabilities and wider tunability.
We only give here a short overview of the theoretical modelling of the transition line and the atomic velocity beam distribution, crucial to estimate systematic effects. 
Indeed, the main systematic effect in our experiment is the second-order Doppler (SOD) effect, which is on the order of 60 kHz for Deuterium atoms, and depends on the velocity distribution of our room-temperature atomic beam. In order to determine this distribution, we follow a method detailed in \cite{Biraben1991}.  It consists in applying a transverse magnetic field in the interaction region, so that an atom moving with velocity $\vec{v}$ experiences a motional electric field $\vec{E} = \vec{v} \times \vec{B}$. The resulting motional Stark shift experienced by the atoms is proportional to the square of their velocity, like the SOD shift. 
For a particular choice of the field intensity, in the neighbouring of a level crossing, it can compensate the second-order Doppler effect
for all atomic velocities in the beam. Actually, taking into account the motional Stark and Zeeman effects, the shift of the SOD effect on 
the transition is only partially reduced, because only some hyperfine $m_F$ sublevels are involved. Therefore, this method is a mean to study the atomic velocity distribution in the beam, by fitting the data with theoretical profiles depending on the applied magnetic field and having as the fit parameters our velocity profile variables. To extract the best parameters of the velocity distribution, we record the transition signal for no applied magnetic field (residual field of 0.04 mT) and for different values of the magnetic field. To avoid bias due to a possible stray electric field, we also reverse the magnetic field direction. More details about the procedure can be found in \cite{Fleurbaey2018}.  

Our velocity distribution model :
\begin{equation}
 f(v,\sigma\, v_0) \sim v^3 \exp{(-\frac{v^2}{2\sigma^2})}\, P(\frac{v}{\sigma}) \,\exp{(-\frac{v_0}{v})}
\end{equation}
is based on the Maxwellian-type distribution of an effusive beam ($\sigma =\sqrt{ \frac{kT}{M}}$, $T$ temperature, $M$ atomic mass) \cite{Fleurbaey2018} and includes the correction $P(\frac{v}{\sigma})$ which describes a depletion of slow atoms due to interactions within the nozzle \cite{Olander70}. It is multiplied by an exponential decay term to model a possible additional depletion of the slow atoms in the effusive beam. This atomic velocity distribution can be fully modelled by the two parameters $\sigma$ and $v_0$ \cite{Galtier2015}. 
The data analysis relies on a theoretical line profile described in \cite{Arnoult2010} which includes the SOD and motional Stark and Zeeman shifts. Thus, we can generate fully derived theoretical line profile, which depends on the velocity distribution parameter $\sigma$ and $v_0$, allowing to extract these parameters with a $\chi^2$ minimization analysis. We finally take into account broadening effects, mainly due to transit time and pressure broadening by convoluting the theoretical line with a Lorentzian function, with its width $\Gamma$ being a free parameter of the numerical fit. \\
This computed theoretical line profile can applied to fit the data and extract the resonance frequency at zero magnetic field $\nu_0$, as illustrated in Fig.\ref{fig3oldfit} a). The derived theoretical line profile $\mathcal{F}^{\sigma,v_0}_{th}$, computed for an atomic velocity distribution with parameters ($\sigma,v_0$) is broaden by convolution with a Lorentzian ($\mathcal{L}$or) curve of Full Width at Half Maximum (FWHM) $\Gamma$, such that the fit model is :

\begin{equation}
\mathcal{F}_{\mathrm{fit},1}(\nu_0,\Gamma) = \mathcal{F}_{\mathrm{th}}^{\sigma,v_0}(\nu_0) \otimes \mathcal{L}or(\Gamma)
\label{eq:FsingleL}
\end{equation}

It is worth noticing that this computed line profile is asymmetric mainly because of the $2^{\mathrm{nd}}$  Doppler order effect, so that the determination of the velocity distribution of the beam is crucial for the proper analysis of our data. 

\subsection{A newly spotted contribution to the fluorescence signal }

\begin{figure*}[!ht]
\centering
\includegraphics[width=0.9\textwidth]{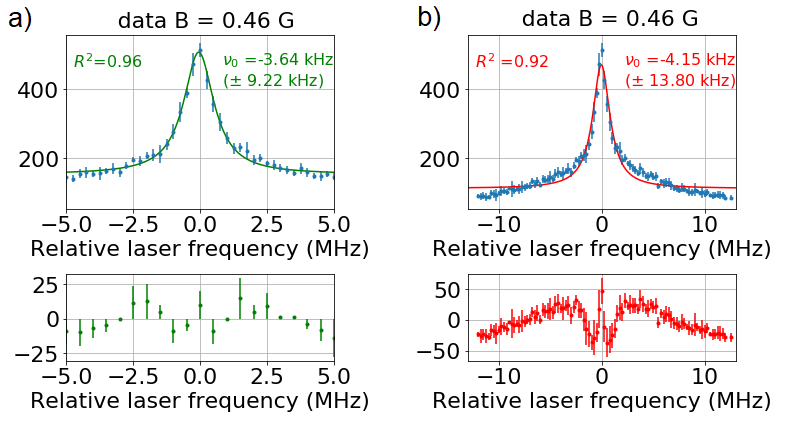}
\caption{A typical fluorescence 1S-3S signal of deuterium atoms, recorded at B~=~0.46~G. a) The averaged counts/s (over 1 run = 10 scans) and the best fit results (green curve) obtained with the fit model of Eq.\ref{eq:FsingleL}, applied on the centre data $[-5,+5]$~MHz range. b) Same data, plotted over the whole $[-12.5,+12.5]$~MHz range, and the best fit results (red curve) with the fit model of Eq.\ref{eq:FsingleL}, to fit he whole data range. The according residuals of the best fit to the data are plotted in the bottom of the figures.}
\label{fig3oldfit}
\end{figure*}

Thanks to the recent improvement on laser stability and tunability explained in Sec. \ref{sec3}, we were able to record a couple runs of data over an extended frequency range, for different magnetic field values. Each run is composed of 10 scans of 72 data points recorded at different laser frequencies. One run takes about 20~min acquisition time.  To extract the resonance frequency $\nu_0$, we then proceed to a python fit, using $\mathcal{F}_{\mathrm{fit},1}$.  
A typical spectrum recorded for $B = 0.46 \, \mathrm{G}$ (residual Earth magnetic field), with a scanning laser frequency range of $[-12.5\,,\,+12.5]$~MHz, is shown in Fig.\ref{fig3oldfit}. In a), we applied the fit model Eq.(\ref{eq:FsingleL}) only to the centre data, restricting the fitted data to $[-5\,,\,+5]$~MHz, to compare to the usual scanning ranges used in the past measurement campaigns. The fit appears to be in good agreement, as the correlation coefficient of the fit compared to the data is $R^2=0.96$, and the fitted uncertainty obtained for a single run is about 10~kHz. This uncertainty is averaging down to a couple of kHz after 1 day of measurements. In Fig.\ref{fig3oldfit} b), we present the result of the fit performed over the whole data, still using the same theoretical line  of Eq.(\ref{eq:FsingleL}). In this case, it appears that the theoretical line profile fails to follow the "tail" of the data:  the fluorescence signal from the hydrogen beam sits on a "pedestal" fluorescence signal that this first model doesn't take into account.

We investigated the origin of this pedestal signal, varying the laser intensity injected into the Fabry-Perot cavity as well as the pressure inside the vacuum chamber, varying the flux of hydrogen in the atomic beam. 
It appears that this signal varies as the square of the laser intensity and decreases when the atomic pressure decreases. We attribute this pedestal signal to a fluorescence 1S-3S signal of the rest-gas hydrogen atoms, that accumulate in the vacuum chamber because of the difficulty of pumping out the hydrogen atoms. The H atoms in the vacuum chamber can cross the laser beam in any direction, interacting with the laser for a shorter period compared to the hydrogen of the atomic beam that leads to a broadening of the transition line compared to the fluorescence of the H beam ones, due to finite transit time (see \cite{PhD_Thomas, PhD_Fleurbaey} for more details on the finite transition time broadening). In order to take into account the contribution of the rest gas H atoms to the total fluorescence signal, we modified our fitting model, adding  to the previous $\mathcal{F}_{\mathrm{fit},1}$ an extra Lorentzian curve to model the pedestal 
signal, with the centre ($\nu_p$) and the FWHM ($\Gamma_p$) as free fit parameters as well:

\begin{equation}
    \mathcal{F}_{\mathrm{fit},2} = \mathcal{F}_{\mathrm{th}}^{\sigma,v_0}(\nu_0) \otimes \mathcal{L}or(\Gamma) + \mathcal{L}or(\nu_p,\Gamma_p) 
    \label{eq:DoubleCurves}
\end{equation}

\begin{figure*}[!ht]
\centering
\includegraphics[width=0.9\textwidth]{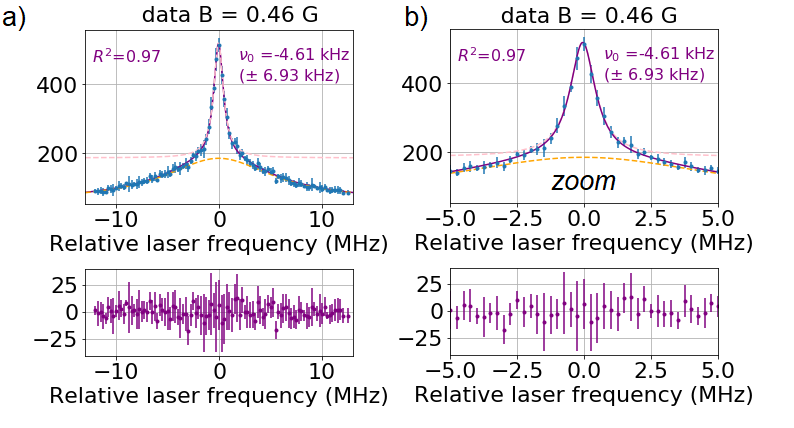}
\caption{ Same fluorescence 1S-3S data of deuterium atoms, as Fig. \ref{fig3oldfit} recorded at B~=~0.46~G. a) The averaged counts/s (over 1 run = 10 scans) and the best fit results (purple line) obtained with the fit model of Eq.(\ref{eq:DoubleCurves}), applied on the whole data $[-12.5,+12.5]$~MHz range. The fitted pedestal signal contribution is plotted in  orange dashed  curve, whereas the pink dashed  line is the fluorescence signal contribution for the atomic beam.  b) Zoom over the $[-5,+5]$~MHz of the fit result of a). The tails appear to be  better fitted than in the case of the model of Eq.(\ref{eq:FsingleL}). The according residuals of the best fit to the data are plotted in the bottom of the figures.}
\label{fig4newfit}
\end{figure*}

Results of the new fit model $\mathcal{F}_{\mathrm{fit},2}$ are displayed in Fig.\ref{fig4newfit}, where the two contributions to the total fit (purple curve), the one coming from the pedestal fluorescence signal (orange dashed curve) and the one from the deuterium atomic beam (pink dashed curve), are displayed. The agreement of this model to the data is better for the tails, as it can be observed on the plotted residuals, and the uncertainty on the centre frequency $\nu_0$ slightly decreases.

We also tried to model the pedestal signal with a broaden theoretical curve $ \mathcal{F}_{\mathrm{th}}^{\sigma_p,v_{0,p}}(\nu_p) \otimes \mathcal{L}or(\Gamma) $, leading to a third fit model $\mathcal{F}_{\mathrm{fit},3}$ :
\begin{equation}
    \mathcal{F}_{\mathrm{fit},3} = \mathcal{F}_{\mathrm{th}}^{\sigma,v_0}(\nu_0) \otimes \mathcal{L}or(\Gamma) + \mathcal{F}_{\mathrm{th}}^{\sigma_p,v_{0,p}}(\nu_p) \otimes \mathcal{L}or(\Gamma_p) 
    \label{eq:DoubleCurvesv2}
\end{equation}

The results with the fit $\mathcal{F}_{\mathrm{fit},3} $ give similar parameters compared to the ones obtained with  $\mathcal{F}_{\mathrm{fit},2} $. This result was expected, since the natural linewidth of the transition (modelled by $\mathcal{F}_{\mathrm{th}}$) is 1~MHz and the pedestal signal is about 8-10~MHz wide. The convolution between the two curves results mostly on a broaden Lorenztian curve. The asymmetry of the fluorescence signal seems to be flatten out by the short transit-time broadening, and within the uncertainty of the fits, we cannot distinguish the two models. 

\subsection{A significant new systematic effect ? }
Once the pedestal signal was revealed, the stringent question was to estimate what could be the maximal contribution of this parasitic signal to the fitted $\nu_0$ centre frequency of the 1S-3S transition. Indeed, the velocity distribution of the hydrogen atoms in the rest gas is most likely different from the one of the effusive atomic beam, since it can be modelled in first approximation as a thermal gas. Within this thermal rest gas, the atoms trajectories are constrained by the geometry of the chamber, with the collimation irises and the collective optics in the way. Overall, the velocity distribution of atoms contributing to this pedestal fluorescence has no reason to be similar to the one of the effusive, collimated, atomic beam, leading to a different $2^{nd}$ order Doppler effect, and thus, to a different asymmetry of the line profile. As a consequence, this effect could potentially shift the apparent centre frequency $\nu_0$, as the signal of interest sits on a wider asymmetrical curve. As a matter of fact,  this effect is quite limited in our data: the fitted  $\nu_0$ values of the centre of 1S-3S transition at zero field, extracted with different fit models and applied either on the centre data (Fig.\ref{fig3oldfit} a) ) or on the whole data range (Fig.\ref{fig3oldfit} b) and Fig.\ref{fig4newfit}), all agree to each other within our current uncertainty. Indeed, the fluorescence signal from the rest gas H atoms is so broaden that it appears flat at the position of the fluorescence peak from the excited atoms from the atomic beam, within the uncertainty of our data. The fluctuations of the intensity signal at constant laser frequency are mainly due to fast instabilities on the enhancement Fabry-Perot cavity lock and the injected 205~nm laser beam intensity.  During the 2020 measurement campaign on deuterium, we recorded only a couple of exploiting runs with extended frequency scanning range for each magnetic field value.  
In order to conclude on the quantitative effect of this pedestal fluorescence signal, we need to accumulate more runs at wider laser frequency ranges. This preliminary study can only state that if any systematic effect exists, it is currently not detectable within our current uncertainty. In plus of accumulating statistics, the intensity fluctuations of the data needs to be reduced, in both short and long timescales. 

\section{Conclusion and perspectives}

In this paper, we first discussed  the recent improvements realised on our apparatus to perform the in-beam two-photon 1S-3S cw spectroscopy of hydrogen and deuterium atoms. Thanks to these upgrades, we then presented a potential new systematic effect that was recently spotted: the presence of a pedestal fluorescence signal that sits below the fluorescence peak coming from the excited atoms in the hydrogen beam. 
The preliminary studies, performed on deuterium during the 2020 measurement campaign, tend to confirm our hypothesis from the origin of this pedestal signal, that comes from the rest gas of deuterium atoms, that are not pumped fast enough and then accumulated; they can get excited by the 205~nm laser beam during a short interaction time while crossing the light. This short excitation period leads to the broadening of the 1S-3S line and the aspect of this broaden pedestal signal of typically 8-10~MHz linewidth. The updated theoretical models (Eqs.\eqref{eq:DoubleCurves} and \eqref{eq:DoubleCurvesv2}) allow to extract the centre frequency $\nu_0$ with a slightly better uncertainty on all the fit parameters than the one obtained with the single theoretical line (Eq.\eqref{eq:FsingleL}) model. But the results of the different fit models are the same within the uncertainties. Thus, even if a more systematic study on this effect needs to be carefully performed and is planned for the next measurement campaign, we can currently state that this effect is not major and most likely cannot explain the discrepancy observed between the measured values of the 1S-3S frequency \cite{Fleurbaey2018}\cite{Grinin2020}.
We are currently finishing some major upgrades on the apparatus, mainly focusing on two points:  the laser system and the vacuum chamber. Concerning the lasers, we decided to change one of our pump laser to generate the 266~nm (VERDI Coherent laser) to a new Y10 Azurlight fiber-amplified 532~nm continuous laser with a claimed bandwidth of about 1~kHz. We have worked on the laser locking system \cite{PhD_Thomas} to be able  to lock all our lasers on the frequency comb, which is itselft stabilised on the reference signal of the 10 MHz from the LNE-SYRTE laboratory \cite{Touahri1997}. This should have an important impact on the reducing the broadening of the line due to the laser bandwidth.  Concerning the vacuum chamber, we opted to change the old oil-pump based vacuum system, to an entirely new H beam and vacuum chamber, pumped with turbo and scroll pumps. We performed mechanical studies to design the chamber such that the vibrations from the pumps would be minimised on the UV mirrors for the enhancement Fabry-Perot cavity (reducing fast fluctuations on the UV beam, ~10~ms-1~s timescale). We will also implement differential pumping stages to try to improve the durability of the UV 205~nm mirror, by injecting a small flux of di-oxygen to refurbish the coating surface, as performed in \cite{Burkley21} and references therein. This would help also for the enhancement cavity long-term stabilisation (about 30~min-1h timescale). All in all, these improvements will allow us to perform soon a new measurement of the 1S-3S transition frequency in hydrogen and deuterium, with careful systematic effects checks and studies; it will bring us to our next goal, the 1S-4S laser spectroscopy at 194~nm, never observed so far and that we can easily address with our laser system with minor upgrades. This work is performed under the following financial assistance award 60NANB18D281 from the US-Department of Commerce, National Institute of Standards and Technology (NIST).
\section{Author contribution statement}
All authors contributed equally to the paper.
\section{Data Availability Statement}
The datasets analysed during the current study are not publicly available since they are part of the whole data campaign of 2020 that is still under analysis, but are available from the corresponding author on reasonable request.

\section{References}

\bibliography{biblioarticle}

\end{document}